\documentclass[aps,twocolumn,superscriptaddress]{revtex4-2}
\usepackage{graphicx,longtable,xfrac}
\usepackage[colorlinks=true,citecolor=blue,linkcolor=blue,breaklinks=true]{hyperref}
\usepackage{amssymb}
\usepackage{graphicx}
\usepackage{amsmath}
\usepackage[version=4]{mhchem}
\usepackage{color}
\usepackage{bm}
\usepackage{indentfirst}

\begin{document}

\title{Fidelity and variability in the interlayer electronic structure of the kagome superconductor \ce{CsV3Sb5}}

\author{Aurland K. Watkins}
\email[]{aurland@ucsb.edu}
\affiliation{Materials Department, University of California, Santa Barbara, California 93106}

\author{Dirk Johrendt}
\affiliation{Department Chemie, Ludwig-Maximilians-Universität München, Munich, Germany}

\author{Vojtech Vlcek}
\affiliation{Department of Chemistry and Biochemistry, University of California, Santa Barbara, California 93106}
\affiliation{Materials Department, University of California, Santa Barbara, California 93106}

\author{Stephen D. Wilson}
\affiliation{Materials Department, University of California, Santa Barbara, California 93106}

\author{Ram Seshadri}
\affiliation{Materials Department, University of California, Santa Barbara, California 93106}
\affiliation{Department of Chemistry and Biochemistry, University of California, Santa Barbara, California 93106}

\begin{abstract}
The \ce{\emph{A}V3Sb5} (\emph{A} = K, Rb, Cs) kagome materials host an interplay of emergent phenomena including superconductivity, charge density wave states, and non-trivial electronic structure topology. The band structures of these materials exhibit a rich variety of features like Dirac crossings, saddle points associated with van Hove singularities, and flat bands prompting significant investigations into the in-plane electronic behavior. However, recent findings including the charge density wave ordering and effects due to pressure or chemical doping point to the importance of understanding interactions between kagome layers. Probing this \emph{c}-axis electronic structure via experimental methods remains challenging due to limitations of the crystals and, therefore, rigorous computational approaches are necessary to study the interlayer interactions. Here we use first-principles approaches to study the electronic structure of \ce{CsV3Sb5} with emphasis on the \emph{k\textsubscript{z}} dispersion. We find that the inclusion of nonlocal and dynamical many-body correlation has a substantial impact on the interlayer band structure. We present new band behavior that additionally supports the integration of symmetry in accurately plotting electronic structures and influences further analysis like the calculation of topological invariants.

\end{abstract}

\maketitle

\section{Introduction}
\par First-principles electronic structure calculations are critical for the identification and characterization of topological materials given the limited and complex experimental tools designed to confirm non-trivial electronic structure topology.\cite{Xiao_TopCalc_2021,Bradlyn_TQC_2017} While bulk band structure features like band crossings or flat bands can point to non-trivial topology, confirmation requires additional techniques. There are three common computational or theoretical methods relying on the bulk band structure to classify topological materials: 1) adiabatically evolving the Hamiltonian to match the band structure of a material with known topology, 2) computing the surface electronic structure to directly detect topological surface states, and 3) calculating a topological invariant. This final method employs the underlying symmetries and features of the bulk band structure to produce an index describing the topology.\cite{Bansil_TBT_2016}
\par A $\mathbb{Z}_{2}$ invariant ($\mathbb{\nu}$) is used to distinguish between trivial ($\mathbb{\nu}$ = 0) and non-trivial ($\mathbb{\nu}$ = 1) time-reversal invariant systems with a bulk electronic gap by specifying whether topological surface states are expected to reside in this  gap.\cite{KaneMele_2005,FuKaneMele_2007,MooreBalents_2007} For systems that additionally possess inversion symmetry, Fu and Kane developed a protocol for calculating $\mathbb{Z}_{2}$ invariants that relies on the parity (inversion) of the Bloch wave functions defined at specific \emph{k}-points that are invariant under time-reversal symmetry. In a 3D Brillouin zone, there are eight of these \emph{k}-points known as time-reversal invariant momenta (TRIM) points. In the Fu-Kane protocol, the product of band parities across all bands below the gap is calculated at individual TRIM points. Subsequently, a product across all the TRIM point products is taken and the $\mathbb{Z}_{2}$ invariant is calculated.\cite{FuKane_Z2_2007}
\par While this method was originally developed for insulators or materials with band gaps, this type of $\mathbb{Z}_{2}$ analysis does not depend on band filling and, therefore, can be applied to metallic systems with partial band occupancy.\cite{Schoop_Au2Pb_Z2_2015,Qian_LaO_Z2_2022,Nayak_LaBi_Z2_2017} However, to extend this calculation to metals, the band structure must have a continuous gap throughout the Brillouin zone (BZ) located near the Fermi level. This requirement ensures that the parity products at each of the TRIM points consistently include the same bands. This method was previously applied to metals featuring kagome nets to support their non-trivial $\mathbb{Z}_{2}$ classification.\cite{Ortiz_KsV3Sb5_Z2_2021,Ortiz_CsV3Sb5_Z2_2020}

\begin{figure}
    \includegraphics[width=0.38\textwidth]{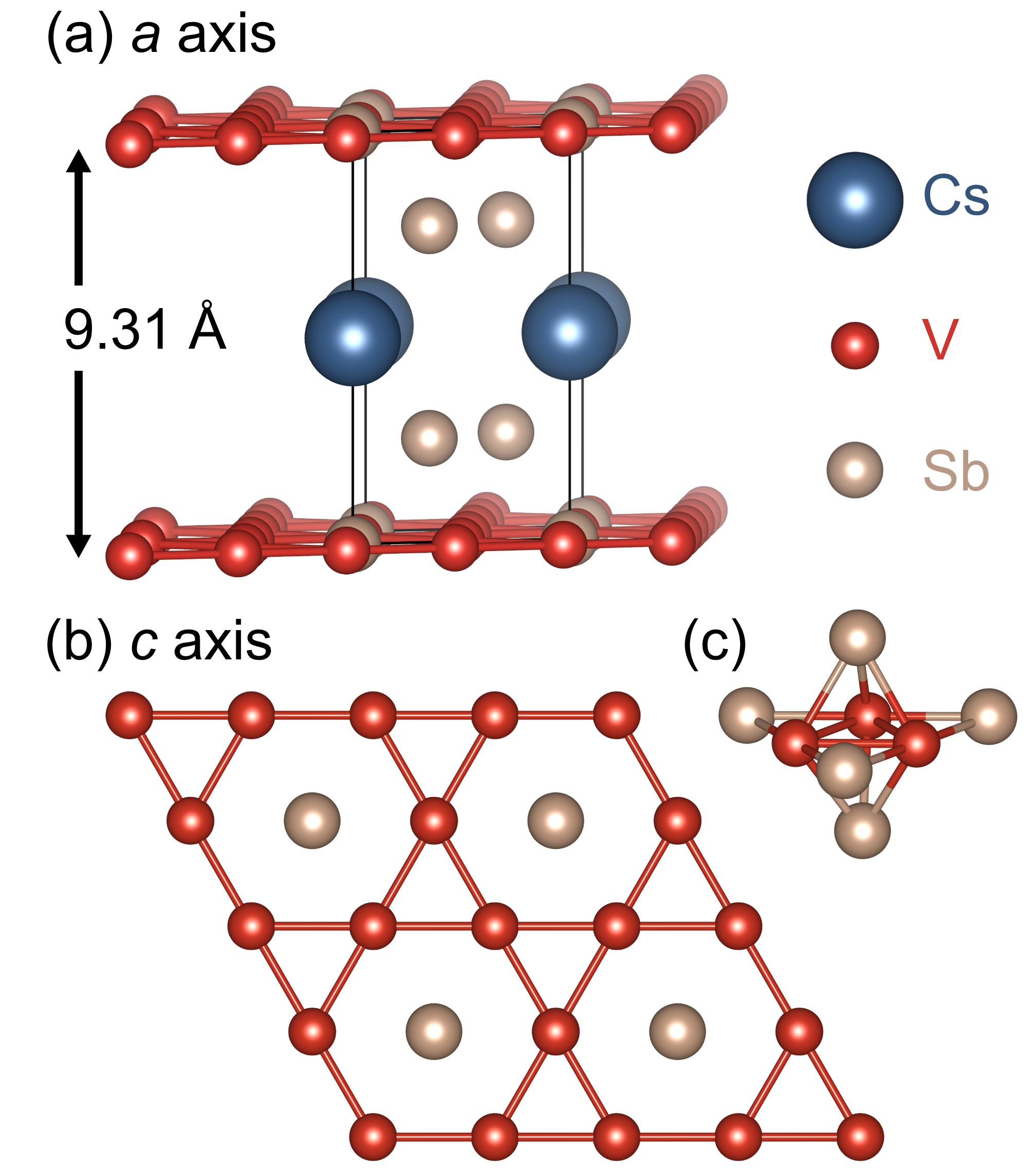}
    \caption{Crystal structure depictions of \ce{CsV3Sb5}. (a) Quasi-2-dimensional structure highlighting separation between kagome sheets. (b) and (c) Coordination environment within and adjacent to kagome layers.}
       \label{fig:Structure}
\end{figure}

\par The kagome net of corner-sharing triangles (Figure \,\ref{fig:Structure}) has drawn significant interest as a platform for a variety of instabilities and emergent phenomena like spin liquid states, superconductivity, charge density waves, and topological surface states.\cite{Yan_kagomeQSL_2011,Ko_kagomesuperconductor_2009,Guo_kagomeTI_2009} Recently, materials within the \emph{A}V\textsubscript{3}Sb\textsubscript{5} (\emph{A} = K, Rb, Cs) kagome family were categorized as $\mathbb{Z}_{2}$ topological metals following the Fu-Kane method.\cite{Ortiz_KsV3Sb5_Z2_2021,Ortiz_CsV3Sb5_Z2_2020} The presence of two continuous gaps throughout the BZ near \emph{E\textsubscript{F}}  allowed for the calculation of multiple $\mathbb{Z}_{2}$ invariants, both of which indicated non-trivial topology suggesting that \ce{CsV3Sb5} exhibits topological surface states analogous to a topological insulator. This original $\mathbb{Z}_{2}$ analysis relied on a band structure that was obtained via a structural relaxation with the PBE functional \cite{Perdew_PBE_1996} and D3 correction with the addition of spin-orbit coupling (SOC) in the self-consistent calculation. The D3 correction is an empirical van der Waals total energy correction that was identified as a necessary parameter to more closely match experimental lattice parameters (specifically along the \emph{c}-axis) during relaxation.\cite{Grimme_D3_2010,Grimme_D3+BJ_2011} 

\par Here, we present electronic structure analysis of \ce{CsV3Sb5}, with special consideration of anisotropic nonlocal correlation along the \emph{k\textsubscript{z}} or inter-kagome-plane direction. Within the framework of density functional theory (DFT),  we find that including certain computational parameters during a structural relaxation gives rise to a symmetry-allowed band crossing previously not identified between $\Gamma$ and A as highlighted in Figure \ref{fig:Bands}. Yet, ultimately, this study suggests that it is critical to include nonlocal and dynamical many-body correlation through the application of the \emph{GW} approximation to obtain accurate interlayer band behavior. The resulting electronic structure shows significant shifts of the bands and the absence of the symmetry-allowed crossing seen in standard DFT band structures. This seemingly minor detail in the electronic structure impacts the calculation of $\mathbb{Z}_{2}$ invariants in this material and points to a larger issue of correct band identification or tracking when plotting band structures.

\section{Methods}
First-principles calculations were performed using the Vienna Ab initio Simulation Package (VASP) version 5.4.4. Pseudopotentials following the projector-augmented wave method were selected with the following valence configurations: Cs (5\emph{s}$^2$5\emph{p}$^6$6\emph{s}$^1$), V (3\emph{s}$^2$3\emph{p}$^6$4\emph{s}$^2$3\emph{d}$^3$), Sb (5\emph{s}$^2$5\emph{p}$^3$). An 11 $\times$  11 $\times$ 5 $\Gamma$-centered \emph{k}-point mesh was automatically generated by VASP and the plane-wave energy cutoff was set to 500 eV. 
Geometric optimization of the experimental structure (ICSD 31841) was performed with all parameters in the description (eg. SOC, D3, TS, etc.) unless otherwise stated and all degrees of freedom (atomic positions and cell shape/volume) were allowed to relax until forces were converged within 10\textsuperscript{$-$7} eV/\AA. All calculations had an energy convergence better than 10\textsuperscript{$-$8} eV. 
A \emph{k}-point path for the band structure was generated using the AFLOW online tool and the density of the path was set 50 \emph{k}-points per path segment (for example between $\Gamma$ and A) to ensure high resolution of band features in the final plot.\cite{Curtarolo_AFLOW_2012} Irreducible representations were obtained using the Irvsp program that interfaces with VASP outputs.\cite{Gao_Irvsp_2021} 
\par \emph{GW} calculations implemented in VASP were performed on the experimental (unrelaxed) structure and employed the PBE electronic structure as a starting point. Both single-shot \emph{G}$_0$\emph{W}$_0$ and partially self-consistent eigenvalue \emph{GW}$_0$ (ev \emph{GW}$_0$) calculations were performed. \emph{GW} potentials were selected based on the VASP recommendations with the following valence configurations: Cs (5\emph{s}$^2$5\emph{p}$^6$6\emph{s}$^1$), V (3\emph{s}$^2$3\emph{p}$^6$4\emph{s}$^2$3\emph{d}$^3$), Sb (4\emph{d}$^{10}$5\emph{s}$^2$5\emph{p}$^3$). The total number of bands was set to 4000 and the number of frequency points was set to 100. Given the high computational cost of \emph{GW} calculations, a smaller 4 $\times$  4 $\times$ 2 \emph{k}-point grid was used. This setup yields quasiparticle energies converged to within 50 meV. For the \emph{GW}$_0$ calculation, seven steps were necessary for convergence, however, all results detailed in this study were taken after ten steps. Since the orbitals are not updated in the ev \emph{GW}$_0$ self-consistency scheme, the interpolation of bands between $\Gamma$ and A in the \emph{GW} output should match that seen in standard DFT even as the eigenvalues are updated. To visualize the resulting band behavior, eigenvalues were matched to the ordering of states in DFT outputs. The interpolation of bands between $\Gamma$ and A was preserved from the unrelaxed PBE band structure, yet it was scaled such that the endpoint energies (at $\Gamma$ and A) correspond to the updated eigenvalues from the \emph{GW} calculations. Further details are provided in the Appendix.

\section{Results}

\begin{figure}
    \includegraphics[width=0.48\textwidth]{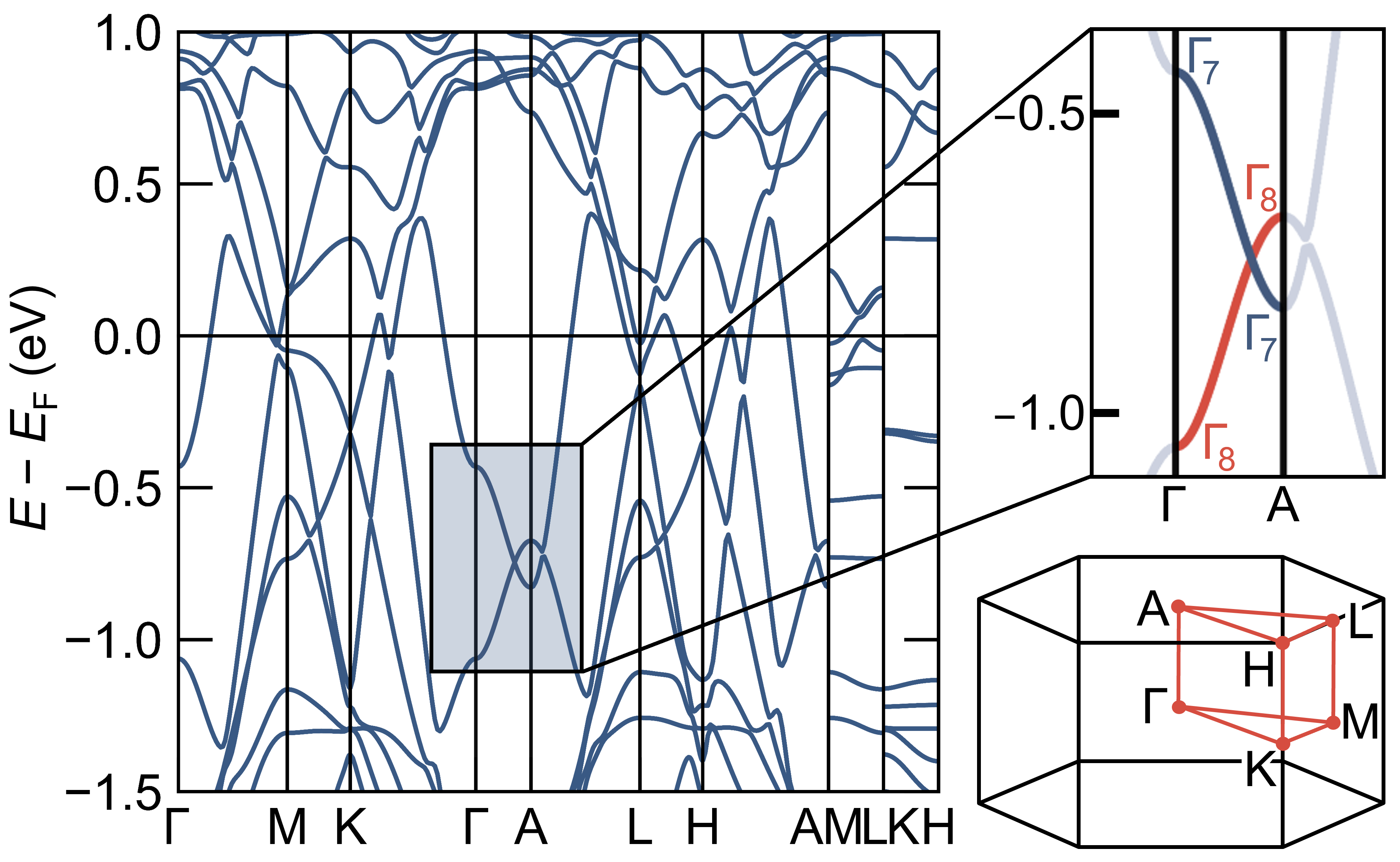}
    \caption{PBE + SOC + D3 electronic structure of \ce{CsV3Sb5} along high-symmetry \emph{k}-point path. A previously unidentified crossing along $\Gamma$-A representing the \emph{k\textsubscript{z}} band dispersion is confirmed by irrep analysis.
    \label{fig:Bands}
    }
\end{figure}

\subsection{General features of the band structure}
\par The electronic structure of \ce{CsV3Sb5} hosts a variety of features associated with unique physical phenomena. Many computational efforts have focused on the Dirac crossings \cite{Hao_Dirac_2022} and van Hove singularities \cite{Hu_CsV3Sb5_TSS_2022} characteristic of the kagome electronic structure. Specifically, there is significant interest in the near-Fermi-level saddle points in the band structure that correspond to van Hove singularities or divergences in the density of states (DOS). Due to the high DOS, interactions can become pronounced when a saddle point becomes populated and can nest across the Fermi surface. In \ce{CsV3Sb5}, multiple saddle points residing at the M-point give rise to competing instabilities depending on band filling.\cite{Kang_Comin_2022,Hu_CsV3Sb5_TSS_2022} 
\par Yet, while the intra-layer interactions draw focus, the interlayer electronic behavior is understudied.\cite{Tsirlin_Sb_2022} Among kagome materials, the 135 family is highly two-dimensional as seen in the stacking of the kagome layers (Figure \,\ref{fig:Structure}). While these materials are not considered traditional 2D materials given the presence of \emph{A}-site ions between the layers, previous electronic structure work concluded that van der Waals interactions were necessary to reproduce the experimental \emph{c}-axis lattice parameter during structural relaxation. These interactions are most relevant along the interlayer or \emph{k\textsubscript{z}} direction represented along the $\Gamma$-A and K-H paths within the Brillouin zone. The near-Fermi level bands along these paths are dominated by Sb p\textsubscript{z}-orbitals unlike the V character bands found elsewhere in the BZ.\cite{Li_orbitalBS_2022}

\subsection{Band crossing along \boldmath{$\Gamma$}-A}

\par When comparing published band structures for this material, there is a large degree of variability with respect to the band behavior between $\Gamma$ and A, specifically around 0.75 eV below the Fermi energy. These band structures can be categorized into three main groups depending on the interaction of the two bands in this energy range: 1) fully gapped bands \cite{Kang_Comin_2022,Oey_CsV3Sb5Sn_2022,Huang_Vergniory_2022}, 2) seemingly touching bands with a minor gap \cite{Ortiz_CsV3Sb5_Z2_2020,Uykur_Tsirlin_2021}, and 3) fully crossed bands with crossings between $\Gamma$ and A and between A and L \cite{Chen_CsV3Sb5_pressure_2021,Yu_pressureZ2_2022}. We similarly obtain these variations of electronic structures depending on the parameters included in the calculation and the methods employed for plotting. To understand the nature of and potentially validate these crossings, the symmetry or the irreducible representations (irreps) of the bands at specific \emph{k}-points needs to be analyzed. When bands possess the same symmetry or irrep, the associated electronic states mix or hybridize and form a gap. Only when bands have different irreps can a symmetry-allowed crossing occur.\cite{Dresselhaus_irreps_2008}
\par Irreps additionally change with the inclusion of SOC since a spinor representation (as opposed to a vector representation) is required to capture the spin symmetry. Within the Brillouin zone of \ce{CsV3Sb5}, most of the paths connecting \emph{k}-points have \emph{C}\textsubscript{2v} symmetry which has only one spinor irrep. Since bands cannot have differing irreps along these paths, there will not be any symmetry-allowed crossings in these areas. However, the paths along $\Gamma$-A and K-H have \emph{C}\textsubscript{6v} symmetry with multiple spinor irreps allowing for crossings. It is important to note that ``symmetry-allowed crossing" does not refer to symmetry-\emph{enforced} crossings arising from nonsymmorphic symmetries. While the crossings of interest are dictated or ``enforced" by the differing symmetries of the bands, we avoid this terminology since it already has a specific definition within topological band theory.\cite{Zhao_enforced_2016,Hirschmann_enforced_2021} 
\par When band symmetry is included in the previously optimized calculations for \ce{CsV3Sb5} (relaxed with PBE + D3 followed by PBE + D3 + SOC ground state calculation in \cite{Ortiz_CsV3Sb5_Z2_2020}), multiple band crossings are observed along $\Gamma$-A and K-H. Since the band behavior between $\Gamma$ and A conflict with previous reports, those crossings are the focus of this study. Based on symmetry considerations, the two bands around 0.75 eV below the Fermi energy cross between $\Gamma$ and A but are gapped beyond A since they possess the same spinor irrep beyond that \emph{k}-point. Although these bands are significantly below the Fermi level, they contribute to the $\mathbb{Z}_{2}$ characterization of this material and they provide insight into the interlayer electronic behavior. 

\par To identify the origin of this crossing, a series of calculations were performed to track the evolution of the band structure and the relaxed physical structure as summarized in Figure \ref{fig:relaxation}. From here on, when referring to electronic structure calculations unless indicated otherwise, this includes a structural relaxation, a self-consistent calculation, followed by band structure and density of states (DOS) calculations in which the Fermi level for the band structure is taken from  the DOS calculation given the more accurate sampling of \emph{k}-points (for more detailed discussion see methods section). Each step in this procedure includes all listed parameters. 
\par When the electronic structure is calculated without the D3 corrections, no band crossing between $\Gamma$ and A is observed in the energy range of interest. Specifically, the PBE band structure shows clearly gapped bands and the PBE + SOC band structure shows a small gap consistent with irrep analysis. These results indicate that some contribution from the D3 van der Waals correction is responsible for the crossing. To deconvolute whether this contribution is structural i.e. from the relaxation or electronic, the same calculations were performed using the experimental structure without any relaxation. The structurally unrelaxed calculations show the band crossing in the absence of D3, suggesting that the contribution from this correction is structural. In other words, the experimental geometry exhibits an electronic structure with the crossing along $\Gamma$-A and the D3 correction maintains this crossing since the relaxed structure most closely matches experimental lattice parameters. 

\subsection{Comparison of band disperson corrections}

\par It is well documented that DFT methods over-delocalize electrons leading to overly disperse bands.\cite{Sholl_DFT_2009, Martin_DFT_2004} These errors are potentially most pronounced for band dispersions along the out-of-plane direction within quasi-two-dimensional materials like the \ce{\emph{A}V3Sb5} kagomes compounds. Within the Brillouin zone (Figure \ref{fig:Bands}), this direction is represented by the $\Gamma$-A, K-H, and M-L paths meaning that these paths are more likely affected by over-dispersion. Since the crossing of interest is along $\Gamma$-A, any correction of the band dispersion (i.e. decreasing the bandwidth) could open a gap between the two bands. 

\begin{figure}
    \includegraphics[width=0.48\textwidth]{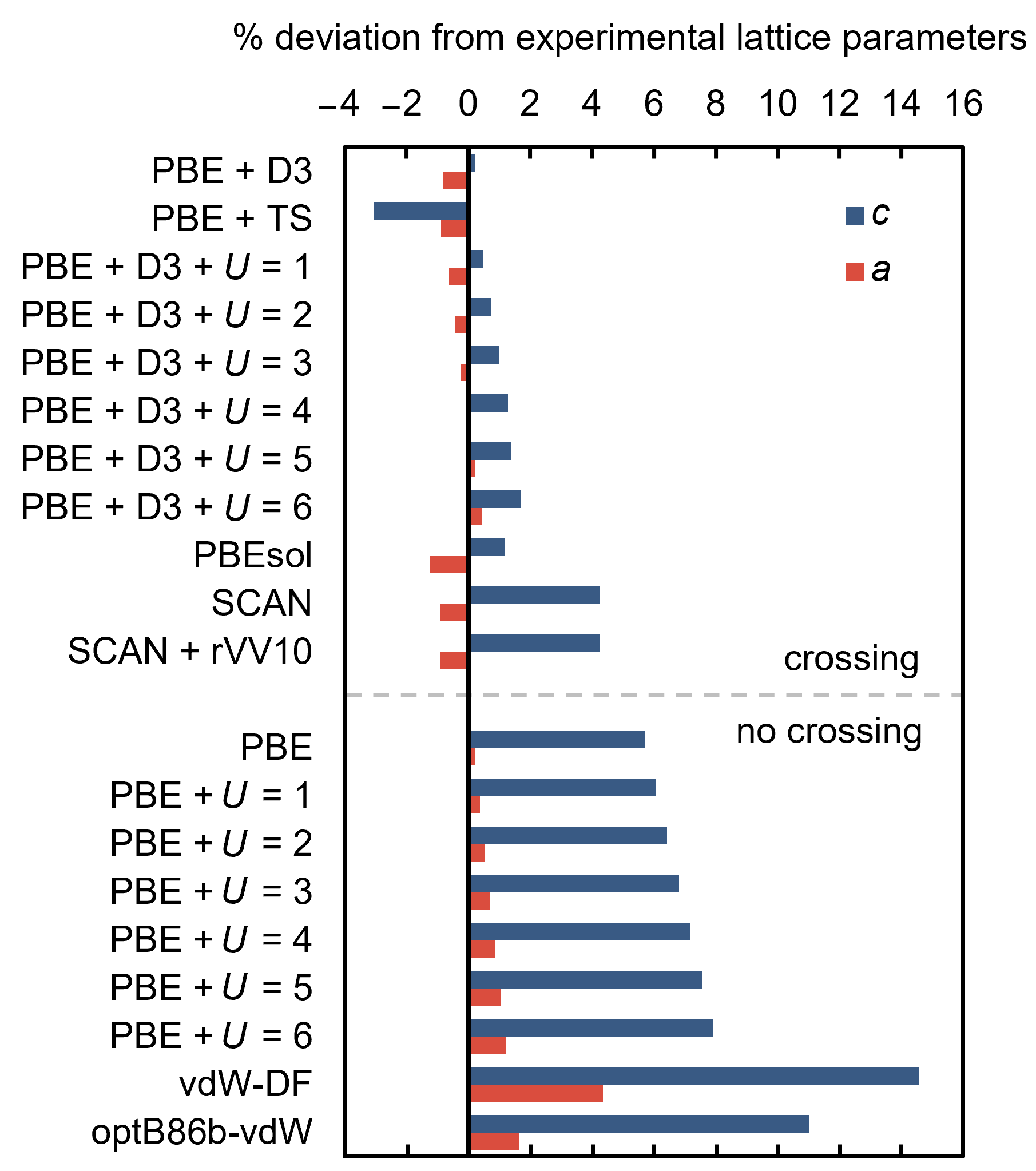}
    \caption{Comparison of computationally-relaxed to experimental lattice parameters. All calculations include SOC. Low deviation from experimental values coincides with the presence of a symmetry-allowed band crossing along $\Gamma$-A.
    \label{fig:relaxation}
    }
\end{figure}

\par To probe the conditions under which a gap might open between the bands and, thereby assess the robustness of this crossing, we compare a variety of commonly applied DFT approaches that address anisotropic nonlocal exchange and correlation at various levels. Again, all parameters or techniques were benchmarked against experimental or unrelaxed results to parse the electronic and structural contributions. Within electronic structure theory, there are a variety of methods to deal with localized electrons including: 1) adding a Hubbard \emph{U} correction to specific orbitals, 2) adding an energetic correction term, 3) switching to a different exchange-correlation functional, and 4) turning to more advanced methods that more accurately treat band dispersions. While the following results do not constitute an exhaustive selection of all methods impacting band dispersion, a representative set of calculations within all of these categories were performed to provide generalized results as summarized in Figure \ref{fig:relaxation}. Methods were selected based on their computational accessibility and the specificity of the correction to the interlayer band dispersion. 

\par Hubbard \emph{U} corrections are commonly used because they provide a targeted method for dealing with electron over-delocalization usually in d- and f-orbitals in a tunable and computationally inexpensive way.\cite{Anisimov_LDA+U_1991,Cococcioni_LDA+U_2005,Kulik_DFT+U_2015,Dudarev_LDA+U_1998} In the context of \ce{CsV3Sb5}, primarily Sb p-orbitals contribute to the near \emph{E\textsubscript{F}} bands at the Brillouin zone center. While p-orbitals typically do not suffer from the same over-delocalization issues in DFT, adding a variable \emph{U} to these orbitals is a valuable exercise for tracking the behavior of the crossing as a function of the bandwidths of the involved bands. As the value of the \emph{U} is increased, the deviations along the \emph{a}- and \emph{c}-axes become progressively more positive signifying an expansion of the unit cell during relaxation. In the presence of the D3 van der Waals correction, calculations with a Hubbard \emph{U} from 1 eV to 6 eV exhibit a crossing. Without the D3 correction, these calculations show fully gapped bands with high lattice parameter deviations along the \emph{c}-axis. 
 
\par Since \ce{CsV3Sb5} is considered a quasi-two-dimensional material given the separation of the kagome layers, van der Waals corrections beyond the D3 term could provide more accurate band dispersions corresponding to localized electrons along the \emph{k\textsubscript{z}} direction. These van der Waals terms are total energy corrections that are added to the calculated Kohn-Sham DFT (KS-DFT) energy and, therefore, do not impact the fundamental electronic structure. The computational efficiency of adding these corrections and the relative success in matching experimental lattice parameters make these terms a popular method for addressing van der Waals systems.\cite{Reckien_DFT-D_2012,Grimme_vdWDFT_2016}  However, moving to less empirical van der Waals corrections like the  Tkatchenko-Sheffler (TS) correction \cite{TkatchenkoScheffler_2009} does not yield an improvement in the relaxed lattice parameters. In fact, the \emph{c}-axis lattice parameter is significantly shorter compared to the experimental value.  

\par Changing the exchange-correlation functional can be a more rigorous method for incorporating dispersion changes within the electronic structure calculation. For example, nonlocal van der Waals density functionals can approximate dispersion interactions by including a nonlocal correlation functional.\cite{RomnPrez_VASPvdW_2009,Dion_vdWDF_2004,Peng_SCAN+rVV10_2016,Klimes_optB86b_2011}  However, for \ce{CsV3Sb5} the selected van der Waals functionals (vdW-DF, optB86b-vdW, SCAN + rVV10) do not perform better than PBE  with the D3 total energy correction in reproducing the experimental lattice parameters. Yet, using PBEsol \cite{Perdew_PBSEsol_2008}, a modification of the PBE functional optimized for extended solids, yielded a relaxed structure with lattice parameters within 2\% deviation from the experimental values without the addition of any explicit van der Waals correction. 
\par All of these corrections and methods were additionally compared to unrelaxed calculations (not shown in Figure \ref{fig:relaxation}) to deconvolute the structural and electronic contributions. Under all tested computational parameters, a crossing is observed between the Sb bands along $\Gamma$-A, pointing to the importance of the structural relaxation in controlling the presence of this crossing. 

\begin{figure*}
    \includegraphics[width=0.85\textwidth]{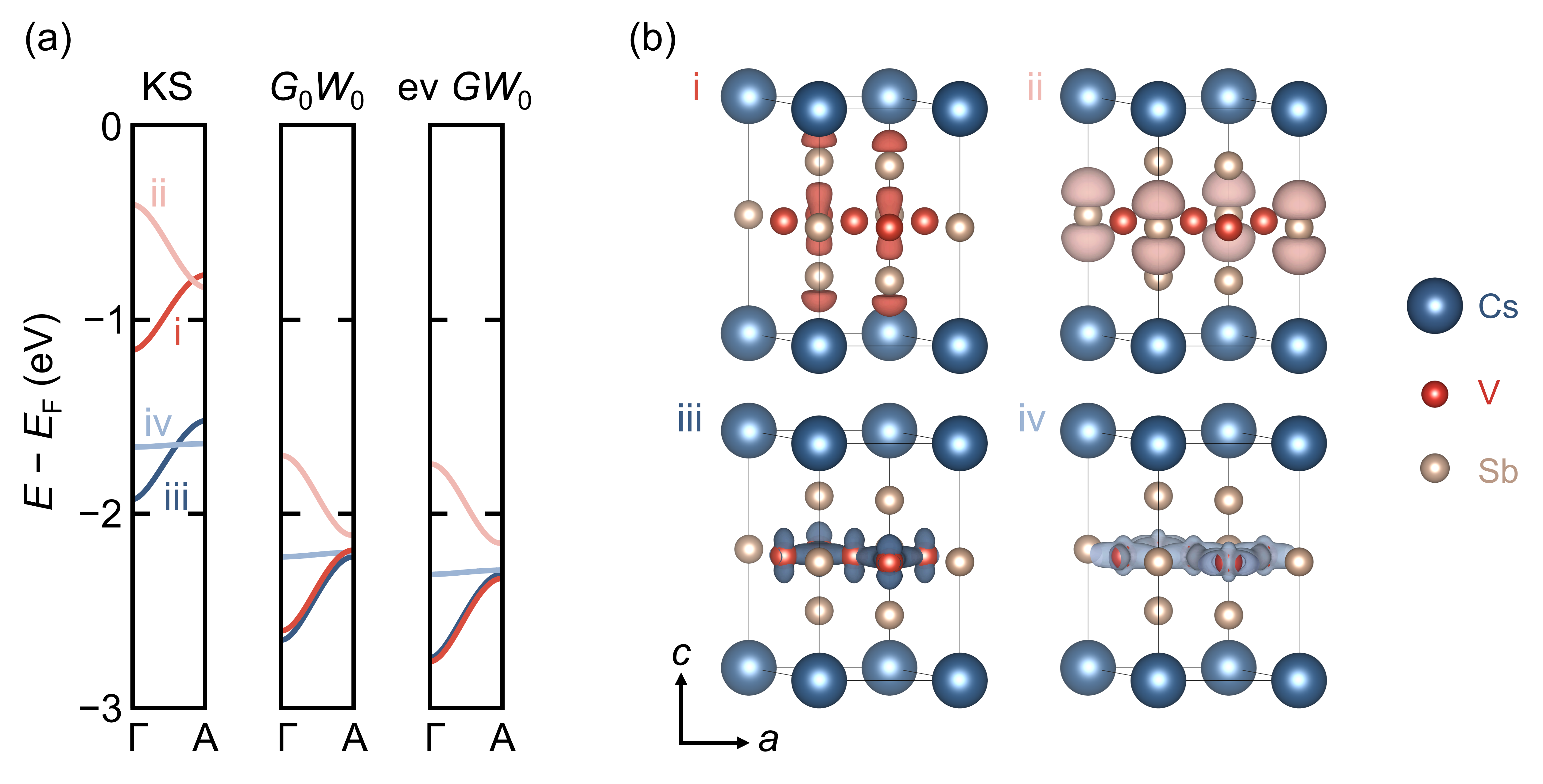}
    \caption{Kohn-Sham (KS) DFT and \emph{GW} interlayer electronic structure behavior. (a) Comparison of KS, single-shot \emph{G}$_0$\emph{W}$_0$, and partially self-consistent ev \emph{G}\emph{W}$_0$ band structures along $\Gamma$-A shows consistent band dispersions yet significant energetic shifts with the opening of the crossing between Sb bands in \emph{GW}. (b) Band-decomposed charge densities corresponding to bands in (a) ordered by degree of interlayer character. i) and ii) represent out-of-kagome-plane and in-kagome-plane Sb p\textsubscript{z} charge density, respectively, while iii) and iv) represent V d charge density with varying interlayer intensity.
    \label{fig:GW}
    }
\end{figure*}

\par Moving outside of the framework of DFT or coupling DFT results with additional methods can also provide a more accurate picture of the electronic structure. For example, using the \emph{GW} approximation for the electronic self-energy can appropriately narrow bands by more accurately capturing electron exchange and correlation, however, at higher computational cost. The \emph{GW} approximation relies on a quasiparticle (QP) picture of the system in which electrons are ``dressed" with positively-charged polarization clouds due to the Coulomb repulsion of other electrons. These quasiparticles, which can be described using single-particle Green's functions (\emph{G}), interact weakly via a dynamically screened Coulomb potential (\emph{W}). Typically, \emph{GW} calculations provide good estimates of the band energies and dispersions when comparing electronic structures from theory and experiment. Since \emph{GW} calculations capture correlations stemming from dynamical and non-local density fluctuations, these calculations are well-suited for systems in which van der Waals forces play a role. For \ce{CsV3Sb5}, these types of interactions likely dominate the interlayer electronic behavior, meaning that the \emph{GW} approximation provides inherent advantages over static mean field methods like KS-DFT. 

\par \emph{GW} calculations were performed on the experimental structure without SOC and compared to a range of DFT results using the LDA, PBE, SCAN, and nonlocal van der Waals density functionals (vdW-DF, optB86b, SCAN + rVV10).\cite{RomnPrez_VASPvdW_2009,Dion_vdWDF_2004,Peng_SCAN+rVV10_2016,Klimes_optB86b_2011,Sun_SCAN_2015} Single-shot \emph{G}$_0$\emph{W}$_0$ and partially self-consistent eigenvalue \emph{GW}$_0$ yield similar band behavior as shown in Figure \ref{fig:GW}, indicating that single shot calculations will give a relatively consistent picture of the band structure. Based on the partially self-consistent eigenvalue \emph{GW}$_0$ results, the band dispersions are fairly consistent with a maximum of 12\% deviation in the dispersions of the Sb bands. A similar investigation of V bands around 1.7 eV below the Fermi level yield a significantly higher deviation of up to 48\%. 
\par While the dispersions of the revelant bands remain relatively consistent when moving to \emph{GW} calculations, all bands experience significant energy shifts and re-ordering especially below the Fermi level as indicated in Figure \ref{fig:GW}. In particular, the two Sb bands shift downward by more than 1 eV, combining these bands with lower-lying less-dispersive V bands that experience less than 1 eV shifts downward. This shift induces a reordering of bands as some V bands now lie between the two Sb bands. Quantifying the quasiparticle energy shifts at $\Gamma$ and A shows larger shifts of around 1.6 eV and 1.3 eV for p\textsubscript{z} bands associated with Sb outside the kagome layer and within the kagome layer, respectively and smaller shifts of 0.80 eV and 0.65 eV for V d\textsubscript{z\textsuperscript{2}} and V d\textsubscript{xy/x\textsuperscript{2}$-$y\textsuperscript{2}} bands, respectively. Orbital contributions are attributed based on band-decomposed charge densities depicted in Figure \ref{fig:GW}b and published electronic structures with orbital projections.\cite{Li_orbitalBS_2022,Tsirlin_Sb_2022} The relative energy shifts of bands with varying degrees of interlayer character ultimately results in bands that are fully gapped between $\Gamma$ and A. 

\section{Discussion}

\par Within standard DFT, various band dispersion correction terms and methods lead to a similar conclusion: when the structure matches or is comparable to the experimental structure, the associated band structure will feature a symmetry-allowed crossing along $\Gamma$-A. This conclusion is evident in calculations where structural relaxations are performed; when there is a small deviation between the relaxed and experimental lattice parameters, a crossing is observed. Unrelaxed calculations reinforce this result by showing a crossing that is robust to a variety of electronic corrections and terms. Additionally, the dispersions of the Sb bands that introduce the crossing are maintained across a variety of functionals and in \emph{GW} calculations. Yet, in moving to quasiparticle energies, significant energetic shifts are observed. These shifts are more pronounced for bands with higher degrees of interlayer orbital character. For example, the purely intra-layer V d\textsubscript{xy/x\textsuperscript{2}$-$y\textsuperscript{2}} bands experience the smallest shift in quasiparticle versus Kohn-Sham energies, whereas the out-of-plane Sb p\textsubscript{z} band experiences the largest shift. Since the in-plane Sb band does not shift as much as the out-of-plane Sb, a gap is expected to open between these bands (note the separation between bands i and ii in Figure \ref{fig:GW}).

 \par Based on DFT calculations, the presence of this symmetry-allowed crossing and its robustness with a variety of band dispersion corrections suggests a reevaluation of the electronic structure of \ce{CsV3Sb5} specifically along the \emph{k\textsubscript{z}} direction. \emph{GW} calculations provide further insights into the interlayer electronic structure, revealing the absence of this crossing due to significant energetic shifts of bands below the Fermi level. Additionally, these results show interlayer interactions mediated through the V and out-of-plane Sb atoms indicated by the almost degenerate i and iii bands shown in Figure \ref{fig:GW}. The hybridization of these states suggests more interlayer phenomena at play than indicated in previous electronic structures.  
 
 \par Given the two-dimensionality of this kagome system, it is particularly important to study the interlayer electronic behavior as has been demonstrated computationally and experimentally with effects due to pressure and chemical doping and charge-density wave ordering.\cite{Oey_CsV3Sb5Sn_2022,Chen_CsV3Sb5_pressure_2021,Yu_pressureZ2_2022,Kautzsch_CDW_2023} All of these phenomena show dramatic changes along the \emph{c}- or \emph{k\textsubscript{z}}-axes pointing to the critical role of understanding interlayer interactions. Rigorous electronic structure calculations capturing the \emph{k\textsubscript{z}} dispersion are further necessary since experimental verification of theses bands via ARPES remains challenging. The highly two-dimensional single crystals make it difficult to obtain a clean surface required for the technique along the relevant axis.
 
\par Additionally, the $\Gamma$-A crossing (or lack thereof) has greater implications in the topological characterization of this system via $\mathbb{Z}_{2}$ invariant calculations. A crossing implies the absence of a continuous gap necessary for Fu-Kane $\mathbb{Z}_{2}$ calculations across a band manifold. In the case of \ce{CsV3Sb5}, there is an additional continuous gap throughout the Brillouin zone that interacts with the Fermi level that yields a non-trivial $\mathbb{Z}_{2}$ invariant. Since the Fu-Kane method involves taking the product of parity eigenvalues, the ordering of bands below the continuous gap does not affect the overall result meaning that the $\mathbb{Z}_{2}$ calculation for this particular gap is unchanged from previous reports even with the band re-ordering along $\Gamma$-A seen in the \emph{GW} results.\cite{Ortiz_CsV3Sb5_Z2_2020} Ultimately, the presence or absence of this symmetry-allowed crossing does not change the non-trivial $\mathbb{Z}_{2}$ classification even though it potentially reduces the number of continuous gaps.  However, in other systems that do not have additional continuous gaps, subtle crossings that are not identified until irreps are considered could have more severe consequences on topological invariant calculations. 

\par This study motivates revisiting the protocol for defining $\mathbb{Z}_{2}$ invariants in metallic systems. Since the Fu-Kane method was originally intended for insulators, the presence of a gap and, subsequently, a separable band manifold is guaranteed by the system. \cite{FuKane_Z2_2007} However, when moving to metals, this condition of separable bands becomes a prerequisite step before continuing with the calculation. From this study of \ce{CsV3Sb5}, we find that visual identification of a near-Fermi level gap that extends throughout the entire Brillouin zone is not always sufficient fulfillment of this prerequisite. Our results highlight the importance of confirming a gap via irreps prior to performing $\mathbb{Z}_{2}$ analysis in metals. 

\par Finally, broader lessons regarding the interpretation of band structures can be learned from this study of \ce{CsV3Sb5}. For example, recognizing that most band structure outputs do not include band symmetries and can, therefore, misidentify ``continuous" bands advocates the integration of irrep analysis when depicting and understanding band structures. This lack of proper band tracking is reinforced if other computational parameters like \emph{k}-point density along the Brillouin zone path are not high enough to resolve small features. These issues are crucial when attempting to calculate topological invariants based on bulk band structures but are also widely applicable to any system in which an accurate picture of the electronic structure is necessary. 

\section{Conclusion}

Using first-principles calculations, we have identified new band behavior along the interlayer direction in \ce{CsV3Sb5}. Given the two-dimensional nature of this system, capturing the electronic structure along the \emph{c}- or \emph{k\textsubscript{z}}-direction is experimentally and computationally difficult. In this study, we survey a variety of computational methods to address nonlocal exchange and correlation at different scales. Density functional theory calculations show a symmetry-allowed crossing along the \emph{k\textsubscript{z}}-dispersion that arises due to the structural contribution of dispersion corrections and methods employed during relaxation to more closely match experimental lattice parameters. This crossing is revealed upon irrep analysis of the bands. Yet, in turning towards dynamical nonlocal correlation via \emph{GW} calculations, this crossing is absent
due to relative energy shifts of bands with varying interlayer character. The resulting \emph{k\textsubscript{z}} electronic structure points to increased interactions occurring out of the kagome plane and serves as a baseline band structure for further studies of inter-kagome-layer phenomena. 

\section{Acknowledgments}
We thank J. W. Harter, L. Balents, M. G. Vergniory, and I. Robredo for helpful discussions and feedback. This work was supported by the National Science Foundation (NSF) through Enabling Quantum Leap: Convergent Accelerated Discovery Foundries for Quantum Materials Science, Engineering and Information (Q-AMASE-i): Quantum Foundry at UC Santa Barbara (DMR 1906325). Additionally, we thank NSF CAREER award DMR 1945098. We also acknowledge the use of computational facilities administered by the Center for Scientific Computing at UC Santa Barbara supported by NSF CNS 1725797 and NSF DMR 2308708. 

\appendix*

\section{Plotting electronic structures}

\par Since typical band structure data include the \emph{k}-points and energies of each band, portions of this data need to be rearranged to reflect the true band identity confirmed by irreps. For example, if two bands cross between \emph{k}-point 1 and \emph{k}-point 2, the output may incorrectly identify all lower energies as a continuous band and similarly all higher energies as another continuous band. In other words, the lower energy band before the crossing will be spuriously connected to the lower energy band after the crossing. However, looking at the irreps at \emph{k}-point 1 and \emph{k}-point 2 will indicate there is a crossing because the irreps switch between the two \emph{k}-points. This weakness in the output can be corrected by identifying the point at which the crossing occurs and reconnecting the bands to match the irrep results. This practice can be seen in Figure \ref{fig:Bands} where the bands between $\Gamma$ and A have been reconnected and colored to match the irreps.

The \emph{GW} band structures in Figure \ref{fig:GW} were constructed based on the unrelaxed PBE band structure. Once the bands were reconnected based on irreps, each band was rescaled and repositioned according to the renormalization computed in the \emph{G}$_0$\emph{W}$_0$ and \emph{GW}$_0$ calculations at $\Gamma$ and at A. As such, bands from the \emph{GW} output may suffer from the same misidentification across crossings, however since only the energies and not the orbitals are updated in the employed partially self-consistent scheme, the \emph{GW} energies can be properly matched based on the ordering of bands from the original PBE band structure.

\bibliography{references}

\end{document}